\documentclass[conference]{IEEEtran}
\IEEEoverridecommandlockouts
\usepackage{cite}
\usepackage{amsmath,amssymb,amsfonts}
\usepackage{algorithmic}
\usepackage{graphicx}
\usepackage{textcomp}
\usepackage{xcolor}

\usepackage{url}
\usepackage[bookmarks=false]{hyperref}
\usepackage{tipa}

\def\BibTeX{{\rm B\kern-.05em{\sc i\kern-.025em b}\kern-.08em
    T\kern-.1667em\lower.7ex\hbox{E}\kern-.125emX}}

\title{A small vocabulary database of ultrasound image sequences of vocal tract dynamics}

\author{
	\IEEEauthorblockN{
	    Margareth Castillo\IEEEauthorrefmark{1},
		Felipe Rubio\IEEEauthorrefmark{1}, 
		Dagoberto Porras\IEEEauthorrefmark{1},
		Sonia H. Contreras-Ortiz\IEEEauthorrefmark{2} and 
		Alexander Sep\'ulveda\IEEEauthorrefmark{1}}
	\IEEEauthorblockA{
		\IEEEauthorrefmark{1}Escuela de Ingenier\'ias El\'ectrica, Electró\'onica y de Telecomunicaciones (E3T), \\
		Universidad Industrial de Santander, Cll 9 Cra 27, Bucaramanga, Colombia \\
		Email: fasepul@uis.edu.co; alexander.sepulveda@gmailcom}
	\IEEEauthorblockA{\IEEEauthorrefmark{2}Electric and Electronic Engineering Programs, Universidad Tecnol\'ogica de Bol\'ivar \\
		Parque Industrial y Tecnoló\'ogico Carlos V\'elez Pombo-Km 1 V\'ia Turbaco, Cartagena, Colombia}
}

\begin{document}

\maketitle

\begin{abstract}
This paper presents a new database consisting of concurrent articulatory and acoustic speech data. The articulatory data correspond to ultrasound videos of the vocal tract dynamics, which allow the visualization of the tongue upper contour during the speech production process. Acoustic data is composed of 30 short sentences that were acquired by a directional cardioid microphone. This database includes data from 17 young subjects (8 male and 9 female) from the Santander region in Colombia, who reported not having any speech pathology. 
\end{abstract}

\begin{IEEEkeywords}
Speech, Articulation, Ultrasound, Tongue.
\end{IEEEkeywords}

\section{Introduction}

The tracking of articulators movement is of significant interest for several disciplines such as speech therapy, language learning and acoustic speech processing. Regarding speech processing, having the articulatory information, in addition to its acoustic cognate, facilitates the study of the dynamics of speech production mechanism and its relation to the corresponding acoustics.

Several systems have been developed in order to acquire the articulators movement directly from human subjects: x-ray cineradiography, x-ray microbeam, ultrasound, electromagnetic articulography (EMA) and magnetic resonance imaging (MRI) systems. Among these, EMA is one of the most popular nowadays. EMA allows the collection of a considerable quantity of data at a relatively high sample rate, however it is able to track only some points of the vocal tract~\cite{Richmond_2002_PhD}. The other techniques above mentioned provide videos of the vocal tract. Images from x-ray cineradiography have been used in~\cite{Maeda_1990}. In spite of its usefulness, x-rays can be dangerous in case of radiation overexposure; therefore, only a very limited set of image sequences are available. In order to reduce overexposure, the x-ray microbeam system was deviced~\cite{Westbury_1994_manual}. It allows the observation of articulators by tracking the position of small gold spheres attached to them~\cite{Xue_1988_conf}. These two systems are not longer used today. MRI is a powerful tool that is capable of obtaining 2D images and videos of the whole vocal tract. However, its use in speech research is restricted due to several reasons: the speaker needs to be accommodated in supine position inside the MRI cabin during acquisition process, MRI video equipment is expensive and, the measured acoustic speech signal results affected. 

Several databases showing the speech production mechanism have been created. For instance, the x-ray cineradiography database presented in \cite{Munhall_1995} offers twenty-five high quality x-ray films (totalling 55 minutes) corresponding to 14 speakers. Additional gathered X-ray resources are presented in \cite{Fabrice_2011}. Regarding X-ray Microbeam technology, the Wisconsin X-ray Microbeam Database \cite{Westbury_1994_manual} contains usable data for  57 different speakers, 32 females and 25 males. On the other hand, the MOCHA-TIMIT \cite{Wrench_MOCHA_2000} is perhaps the best-known among EMA databases and contains usable articulatory and acoustic data of two speakers (1 male and 1 famale) uttering 460 TIMIT short sentences. TORGO database \cite{Rudzicz_2010}, which is for dysarthric speech, also incorporates EMA technology. In case of MRI, UCS-TIMIT database \cite{Narayanan_2014} contains MRI video data parallel with the acoustic signal corresponding to 10 speakers (5 male and 5 female) uttering 460 TIMIT sentences, it also incorporates EMA data for 5 speakers. 

Ultrasound imaging is another method to obtain vocal tract videos. In ~\cite{Csapo2017c} it is shown an acquisition setup for the recording of tongue movement in midsagittal orientation using and ultrasound scanner. The procedure is carried out for one female Hungarian subject with normal speaking abilities uttering 473 sentences. In contrast to other techniques above mentioned, ultrasound is safe, fast, cost-effective, and it does not affect the acoustic signal. These advantages makes ultrasound an attractive technique for the development of new speech related devices ~\cite{qin2008predicting, preston2017ultrasound, csapo2017dnn, TamasXu2017}. Motivated for all these advantages, in this work we present a new database containing vocal tract ultrasound image sequences, in parallel with their corresponding acoustic speech signals, for 17 Spanish speakers from the Santander region in Colombia.

\section{Data collection}
\label{data_collection}

\begin{figure}[ht]
	\centering
	\includegraphics[width = 8.7cm]{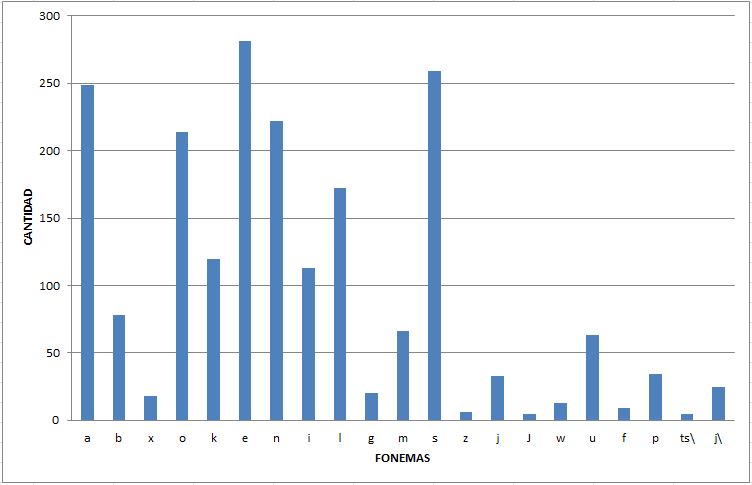}
	\caption{Histogram of phonemes included in the speech corpus.}
	\label{fig:histograma_fonemas}
\end{figure}

\paragraph*{Subjets}
The dataset presented in this work was collected in a room of the \textit{Universidad Industrial de Santander} (www.uis.edu.co). The dataset contains 17 speakers (9 female and 8 male) labeled as {f01, f02, ..., f09; m01, m02, ..., m08}, whose age is in the range of 20-24 years old. All the participants were born in the Santander region of Colombia. Each speaker was asked to utter a set of 30 sentences in Spanish (Colombian Spanish). Figure (\ref{fig:histograma_fonemas} ) shows the phoneme distribution of the sentences belonging to the corpus database.

\paragraph*{Data acquisition}

\begin{figure}[ht]
	\centering
	\includegraphics[width = 6cm]{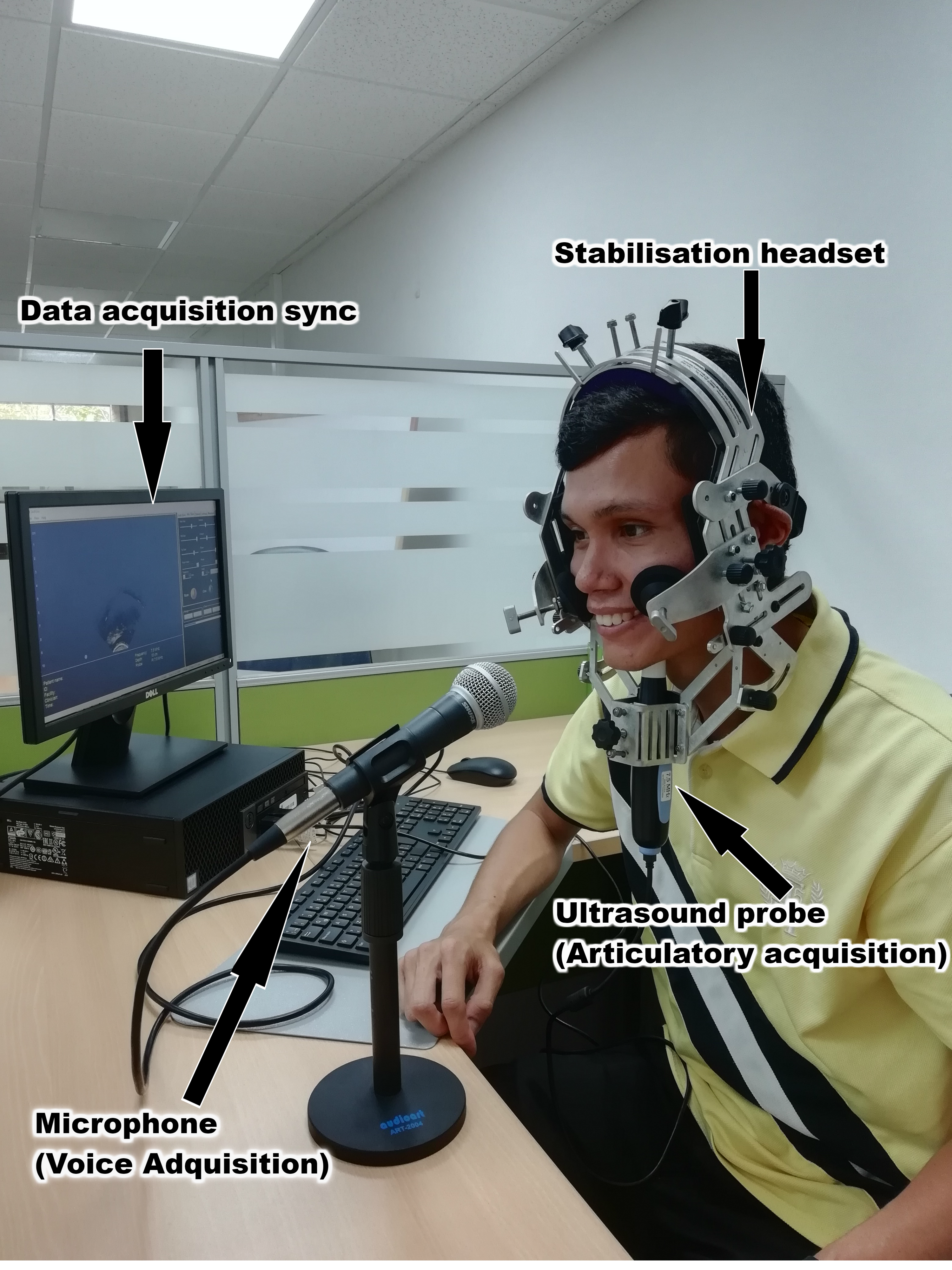}
	\caption{Arrangement of the hardware used for the collection of samples.}
	\label{fig:US_con_Casco}
\end{figure}

The following data was recorded in parallel. First, the acoustic signal was acquired at a sampling frequency of 44000 Hz by means of a \textit{Shure SM58} microphone (http://www.shure.com/americas/products/microphones/sm/- sm58-vocal-microphone) which was located at a distance of about 5 cm from the mouth of the speaker. We used this distance in order to reduce the noise that a person generates when breathing. Second, ultrasound video of tongue's movement in the midsagittal plane was acquired using the PI 7.5 MHz 12-15 fps ultrasound system (\textit {Speech Language Pathology ultrasound system}, http://seemore.ca/portable-ultrasound-products/pi-7-5-mhz-speech-language-pathology-99-5544-can/). Previous to the transductor positioning, conductive gel was applied on the chin to obtain a better image of the tongue contour. In order to avoid variability caused by the head movement and probe positioning, the probe was attached to a stabilizing helmet~\cite{scobbie2008head, HOCUS_Whalen2005} manufactured by \textit{Articulate Instruments}, as shown in the figure~\ref{fig:US_con_Casco}. This allows freedom for the speaker and also guarantees uniformity when recording the data. 

A Python script was developed to synchronize the triggering of acoustic and ultrasound devices for data acquisition. The script uses the \textit{pyaudio} library to collect the audio signal from the microphone. The ultrasound images were acquired using the software associated with the ultrasound probe (\textit{Seemore}), which allows to visualize the ultrasound images in real time and controls the recording of the sequences of images. The audio signals were stored in \textit{.wav} format and the ultrasound videos were saved as sequences of 2D grayscale images (800x600 pixels) in \textit{.jpg} format. Approximately, 340 seconds were recorded for each speaker, corresponding to at least 4000 images per speaker. Table ~\ref{tab: data set} shows the number of images per speaker, totaling 101663 images.

One characteristic of the ultrasound probe is its reduced sampling rate, but articulators, as many biological systems, tend to move slowly. The temporal resolution of articulatory movements is analyzed in several works such as ~\cite{FontecaveJallon2009, Fontecave2009, Ghosh2010}. The authors in \cite{Ghosh2010} quantify the effective maximum frequency content of slowly articulatory varying signals by calculating the frequency $f_c$ below which a certain percentage (say 95\%) of the total energy is contained. They estimated its corresponding mean $\mu_{f_c}$ and standard deviation $\sigma_{f_c}$ values for the speakers in the MOCHA-TIMIT database. From the data reported in \cite{Ghosh2010}, and assuming a normal distribution for those fastest articulators, it can be inferred that there is a probability of 0.977 $P(f_c \leq \mu_{f_c} + 2\sigma_{f_c} )$ that 95\% of the total energy is located below 6.61 Hz and 6.87 Hz for the case of male and female subjects, respectively. On the other hand, the authors in ~\cite{FontecaveJallon2009} observed a low-pass distribution of articulatory information; and in consequence, they fixed a cut-off frequency of 6 Hz for their video recordings. In addition, it is shown in ~\cite {Lofqvist2007} that when speaking in a cadenced way the bandwidth of tongue movement is reduced. For the recording process of the present dataset, words whose pronunciation involves the alveolar trill \textipa{r} (as in Spanish word \textit{ca\textbf{rr}o}) and the alveolar tap \textipa{R} (as in \textit{ca\textbf{r}o}) are excluded from the corpus. Videos belonging to the dataset can be downloaded from this 
\href{https://figshare.com/collections/Vocal_Tract_Ultrasound_Videos_with_Acoustic_Utterances/5260913/1}{FigShare link} \url{https://figshare.com/collections/Vocal_Tract_Ultrasound_Videos_with_Acoustic_Utterances/5260913/1}.

\begin{table}[!hbt]
\begin{center}
\begin{tabular}{|l|r|r|}
\hline
speaker label & \# Sessions & \# Images \\
\hline
f01 & 30 & 7132 \\
\hline
f02 & 30 & 7296 \\
\hline
f03 & 30 & 7522 \\
\hline
f04 & 30 & 7352 \\
\hline
f05 & 30 & 7236 \\
\hline
f06 & 30 & 6958 \\
\hline
f07 & 30 & 7558\\
\hline
f08 & 30 & 7131 \\
\hline
f09 & 30 & 4219 \\
\hline
m01 & 30 & 6813 \\
\hline
m02 & 30 & 7271 \\
\hline
m03 & 30 & 7375 \\
\hline
m04 & 30 & 7193 \\
\hline
m05 & 15 & 3161 \\
\hline
m06 & 20 & 4176 \\
\hline
\hline
m07 & 29 & 4587 \\
\hline
m08 & 30 & 5157 \\
\hline
 & \textbf{Total} & \textbf{101663} \\
\hline
\end{tabular}
\caption{Ultrasound Tongue Images for each speaker labeled in the data set collected.}
\label{tab: data set}
\end{center}
\end{table}

\section{Labeling}

\begin{figure}[ht]
	\centering
	\includegraphics[width = 7cm]{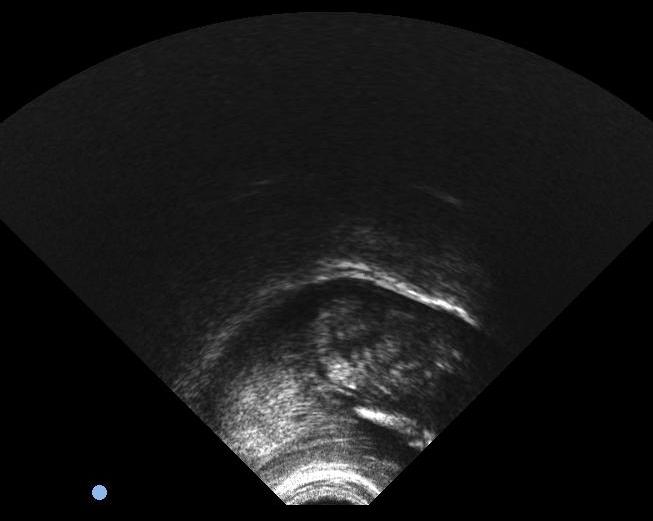}
	\caption{One 2D Ultrasound Tongue Image recorded.}
	\label{fig:US_image}
\end{figure}

\begin{figure}[ht]
	\centering
	\includegraphics[width = 7cm]{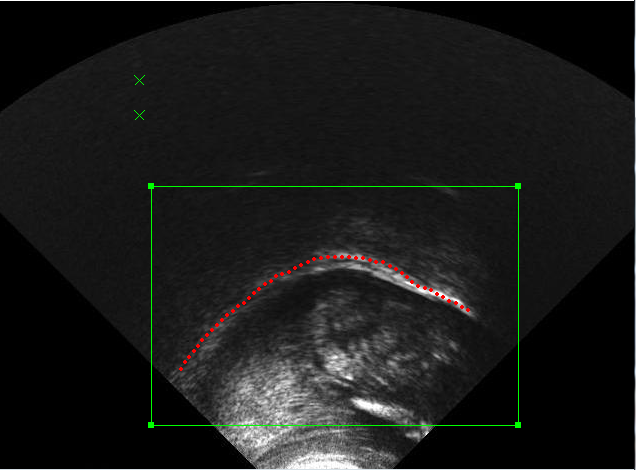}
	\caption{Tongue Contour Extraction from an Ultrasound Tongue Image.}
	\label{fig:Contorno_con_US}
\end{figure}

\paragraph{Tongue contour detection}
A sequence 2D Gray-scale images was obtained from the ultrasound system. One ultrasound image is shown in the figure~\ref{fig:US_image}. The most relevant information shown in the image is the tongue contour. Therefore, it was necessary to process the images for extraction of the tongue contour from each ultrasound image. For this purpose, we used the software EdgeTrak~\cite{li2005automatic}. The procedure is carried out as follows: first, each ultrasound image is loaded into the software, then the area of interest is adjusted (where the contour of the tongue is located). In figure~\ref{fig:Contorno_con_US} is shown the area of interest as the green window. Then the contour must be initialized with 6 points along the tongue and finally the software applies an algorithm based on active contours~\cite{kass1988snakes} which allows to obtain the complete contour. In this final step and similar to~\cite{qin2008predicting}, we observed that in practice the algorithm lost the track of the tongue frequently. For that reason, it was necessary to verify frame by frame visually, and when necessary, to correct the contours manually. Thus, we adopted a semi-automatic approach. This process is time-consuming and it is not commonly carried out in speech ultrasound databases. In the figure~\ref{fig:Contorno_con_US} it can be seen an example of the contour extracted from an ultrasound image. Each tongue contour is exported in a text file, which contains a matrix with the contour coordinates ($ x_m, y_m $) and where $ m = \{1,2, ..., 50 \} $ correspond the number of points extracted.

\begin{figure}[ht]
	\centering
	\includegraphics[width = 8.7cm]{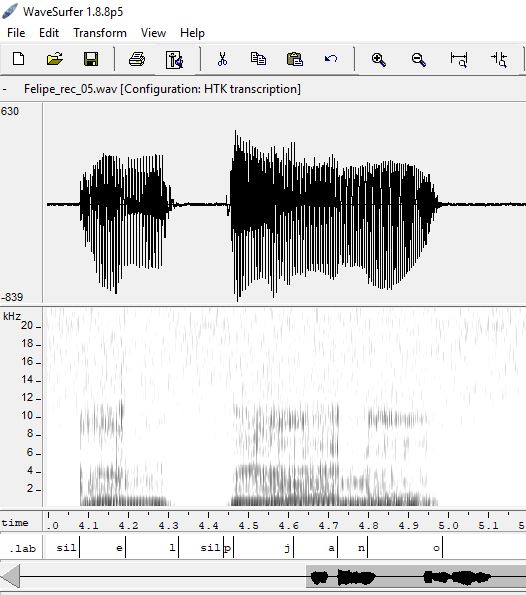}
	\caption{Annotation procedure by using \textit{WaveSurfer}.}
	\label{fig:Annotation_WaveSurfer}
\end{figure}

\paragraph{Annotations}
In order to increase the database usefulness, a segmentation process on the acoustic signals was performed. To do so, the speech signal was segmented manually by using the software \textit{WaveSurfer}. Each audio was loaded and heard, and the spectrograms and waveforms were observed. The silence segments were detected and segmented, and the resulting pieces were segmented into phonemes. The results were saved in a .lab file. The process was performed for each of the 30 sentences uttered by 5 speakers. These results were used to create a dictionary that can be used in the future to create an automatic segmentation process based on HTK, but this is an ongoing process. Figure (\ref{fig:Annotation_WaveSurfer}) shows an example of the segmentation process.

\section{Possible applications}

\subsection{Evaluation of speckle noise reduction algorithms.}

The signal-to-noise ratio of the ultrasound image is quite low \cite{Kele_2016}; and, a considerable part of the noise is speckle. Speckle is a granular 'noise' that inherently exists and degrades the images by concealing fine structures. For the sake of reducing this noise and its influence, anisotropic diffusion filtering can be applied. The SRAD algorithm proposed in \cite{Yongjian_2002} was tested \cite{Janna_2017}, which is a filter that not only preserves the edges, but also improves them. The set of images used in \cite{Janna_2017} correspond to 75 vocal tract ultrasound images from 5 speakers, 15 images per speaker, obtained by using the ultrasound PI 7.5 MHz probe described in \ref{data_collection}.

\begin{figure}[ht]
	\centering
	\includegraphics[width = 7cm]{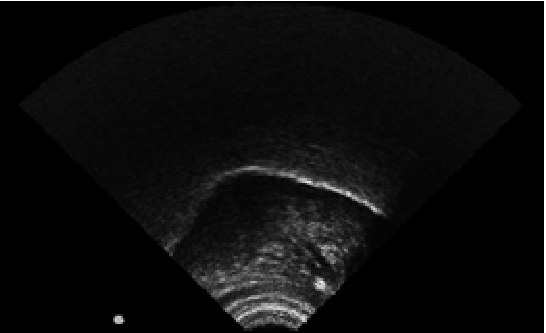}
	\caption{Original ultrasound image.}
	\label{fig:img_original}
\end{figure}

\begin{figure}[ht]
	\centering
	\includegraphics[width = 7cm]{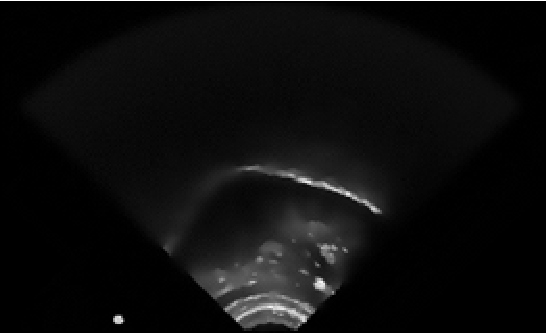}
	\caption{Filtered ultrasound image by using DFSrad algorithm.}
	\label{fig:img_filtered}
\end{figure}

In that work, signal-to-noise ratio (SNR) was used to measure the improvement of the SRAD filter. SNR is defined as the ratio of the average value $\bar{x}$ to the standard deviation of the image in homogeneous regions $\sigma_x$ ($SNR = \frac{\bar{x}}{\sigma_x}$). The percentage improvement over all images was calculated, obtaining a value of 13\%. This result is similar to those obtained in \cite{Sonia_2014}. The effect of the anisotropic diffusion filter can be observed in figures (\ref{fig:img_original}) and (\ref{fig:img_filtered}).


\subsection{Acoustic-to-articulatory inversion}

Recently, there has been a significant interest on acoustic-to-articulatory inversion (AAI), because it could improve the performance of several tasks such as speech recognition~\cite{king2007speech}, speech synthesis~\cite{ling2009integrating}, speaker verification~\cite{li2016speaker}, and talking heads~\cite{wang2010synthesizing}. The AAI methods estimate articulatory movements from the acoustic speech signal, whose data-driven approaches require parallel acoustic-articulatory data for training the AAI model~\cite{Sepulveda2013}. Commonly, the ElectroMagnetic Articulography (EMA) method is used, but, ultrasound Tongue Imaging (UTI) is a technique of higher cost-benefit, thus larger databases could be collected at lower costs. In fact, phonetic research has employed 2D ultrasound for a number of years for investigating tongue movements during speech \cite{Stone2005a, Csapo2015a}.

\subsection{Speech production analysis for improving speaker recognition systems}
Speaker information exists in the articulatory trajectories, thus, if they were correctly estimated they could be applied to speaker recognition systems. In \cite{ghosh2011automatic}, estimated articulators' positions from the acoustic signals are used to improve the performance of speaker verification systems. In addition, the authors in ~\cite{LI2016196} show that augmenting the MFCCs with features obtained from subject-independent acoustic-to-articulatory inversion significantly improves the performance of speaker verification systems. Despite some previous works, assessing the potential usefulness of articulatory features for speaker recognition applications is still an open issue. Regarding this issue, having the manually segmented tongue contours would be very useful.

\section{Conclusions}
This work presents the development of a database of tongue ultrasound images and speech signals. The image and signal quality is good, which allows manual and semi-automatic segmentation. Future work includes the extension of the database with other subjects, using a higher sampling frequency, and the development of automatic algorithms for speech and image segmentation.     


\bibliographystyle{unsrt}
\bibliography{paper_STSIVA.bib}

\end{document}